\documentclass[amsmath,amssymb,aps,prd]{revtex4-2}
\usepackage{amsmath}
\usepackage{ulem}
\usepackage{color}
\usepackage{graphicx}
\usepackage{epsfig}
\usepackage{subfig}
\usepackage{bm}
\usepackage{hyperref}
\usepackage{natbib}
\usepackage{pgfplots,mathtools}
\usepackage{braket}
\usepackage{slashed}
\usepackage[compat=1.0.0]{tikz-feynman}
\usepackage{physics}
\usepackage{xfrac}
\usepackage{algorithm}
\newcommand{\be}{\begin{equation}}
\newcommand{\ee}{\end{equation}}
\newcommand{\bea}{\begin{eqnarray}}
\newcommand{\eea}{\end{eqnarray}}
\newcommand{\ba}[1]{\begin{array}{#1}}
	\newcommand{\ea}{\end{array}}
\newcommand{\nn}{\nonumber}

\newcommand{\ep}{\epsilon}
\newcommand{\om}{\omega} 
\newcommand{\Om}{\Omega}

\begin{document}
\title{Effect of Coriolis force on the electrical conductivity of quark matter: a non-relativistic description}
\author{Ashutosh Dwibedi$^1$, Cho Win Aung$^1$, Jayanta Dey$^2$\\
Sabyasachi Ghosh$^1$}
        \affiliation{$^1$Department of Physics, Indian Institute of Technology Bhilai, Kutelabhata, Durg 491001, India}
	\affiliation{$^2$Department of Physics, Indian Institute of Technology Indore, Simrol, Indore 453552, India}
        \begin{abstract}
Rotating quarks and hadronic systems, produced in peripheral heavy ion collisions, can experience
Coriolis force and other forces due to rotational motion. Considering only the effect of Coriolis force, we have calculated the electrical conductivity for non-relativistic rotating matter using the relaxation time approximation under the Boltzmann transport equation. 
A similarity in mathematical calculations of electrical conductivity at finite rotation and finite magnetic fields is exposed,
where an equivalence role between Coriolis force on massive particle's motion and Lorentz force on charged particle's motion
is noticed. As the beginning level step, we consider only the Coriolis force in the non-relativistic formalism, which will be extended in the future
towards the relativistic case, and to adopt other forces for a more realistic description of the rotating quark and hadronic system.
%
         \end{abstract}
\maketitle
\section{Introduction} 
During non-central heavy ion collisions (HICs), a substantial amount of orbital angular momentum (OAM) is expected to be produced.
Depending on the system size, impact parameter, and collision energy, this initial OAM can vary from $10^3\hbar$ to $10^7\hbar$~\cite{STAR:2017ckg,Liang:2004ph,Becattini:2007sr}. Some fraction of this initial OAM gets transferred to the resulting plasma phase. The initial velocity profile of the resulting quark-fluid encodes this OAM in the form of local vorticity ~\cite{Becattini:2007sr}. Since large OAM can affect the system, the impact of this initial OAM on heavy ion phenomenology, like spin-polarization, chirality, etc., becomes a contemporary matter of interest in the community. Refs.~\cite{becattini2007microcanonical, Becattini:2007sr, BECATTINI20082452, BECATTINI20101566, Becattini:2013fla, Becattini:2013vja, Becattini2015, Becattini:2021suc} have used the global thermal equilibrium under rotation to describe the thermodynamics of the medium and spin-polarizations of final state particles. Whereas Refs.~\cite{Wang_2012, Wang_2013,Wang_2016, Wang_2017,GAO2015542,Yang2017,Gao2017,Gao2018,Liao2018,Gao2019,Gao_2019,Hattori2019,Zhuang_2019M,Rischke2019,yang2020effective,Rischke2021w,Rischke_2021Nk} have used the Wigner function formalism to explain the polarizations and chiral effects associated with the system. On the other hand, the Refs.~\cite{Betz:2007kg, Liang:2004ph, LIANG200520, Wang_2008,chen2009general, XuGuangHuang2011} have used the spin-orbit coupling under strong interactions to obtain the polarizations observed in HICs. More recently, the Refs.~\cite{Becattini2011,Florkowski2018,Florkowski_2018,Florkowski:2018fap,Florkowski:2018ahw,BECATTINI2019419,Florkowski:2019qdp,BHADURY2021,Bhadury_2021,daher2022equivalence,Bhadury2022} have developed a new framework called spin-hydrodynamics by including spin tensor. Spin-polarization and vorticity, the two important quantities in HICs, have been analyzed in various Refs.~\cite{XuGuangHuang2016,Jiang2016,Pang:2016igs,Li:2017slc,Xia:2018tes,XuGuangHuang2019,XuGuangHuang:2019xyr,Wu:2019yiz,XuGuangHuang:2020dtn,Fu:2021pok,XuGuangHuang_2022,XuGuangHuang2022} by using hydrodynamic and transport models. Thermodynamics of the hadronic medium under rotation and the hydrodynamic evolution of QGP in the presence of both the magnetic field and vorticity have also been studied in Refs.~\cite{Pradhan:2023rvf, Sahoo:2023xnu}. 

There is an equivalence and similarity between magnetic field and rotation. Firstly, both are produced in peripheral HIC. The magnetic field can change the motion of charged particles via Lorentz force, while rotation or vorticity do the same for any massive particles via Coriolis force. The equivalence between Lorentz force and Coriolis force has been nicely discussed in the Refs.~\cite{J_Sivardiere_1983, Johnson2000-px,Sakurai1980}. The momentum and spin of the medium constituents (quarks and gluons) produced in HICs would be affected by both rotation and magnetic field. The first quantity(momentum) will be deflected in the presence of rotation via Coriolis force and in the presence of a magnetic field via Lorentz force. On the other hand, the second quantity (spin) in the presence of a magnetic field and/or rotation will be affected by a different mechanism which is linked with the polarization phenomenology. Along with polarization, magnetic fields cause a variety of phenomena, including chiral magnetic effect, magnetic and inverse magnetic catalysis~\cite{FUKUSHIMA2019167,Kharzeev:2015znc,Andersen:2021lnk,Li:2020dwr}. Furthermore, it is evident that the magnetic field has sizeable effects on the observable, elliptic flow~\cite{Roy:19,Inghirami:2019mkc,sun:19}. The electrical conductivity of the medium is the critical property determining the space-time evolution of the magnetic field in the medium. In this paper, we will show how the presence of rotation can affect the electrical conductivity of a medium, contributing significantly to HICs. Similar to the magnetic field case, rotation causes anisotropy in the transport coefficients; however, how both of them (magnetic field and rotation) combinedly affects the transport property requires further investigation.

In the present article, we will focus on the calculation of the electrical conductivity of the medium due to the presence of the Coriolis force. There may be other pseudo-forces that may or may not be equally important to consider in our description; as a first work in this direction, we have focused only on the effect of the Coriolis force; however, other pseudo-forces may have a significant effect that needs to be estimated in future. Moreover, for simplicity, we will take a non-relativistic approach with the future aim of developing a relativistic framework to calculate the conductivity of the rotating medium.
In recent times, Refs.~\cite{Dey:2019vkn,Ghosh:2019ubc,Dey:2020awu,Kalikotay:2020snc,Dey:2021fbo,Satapathy:2021cjp,Das:2019wjg,Das:2019ppb,Dey:2019axu,Chatterjee:2019nld,Hattori:2016cnt,Hattori:2016lqx,Satapathy:2021wex} have thoroughly studied the electrical conductivity of a medium in the presence of a finite magnetic field. By linking the similar nature of the Coriolis force with the Lorentz force, the present article has studied the problem- the effect of the Coriolis force on the electrical conductivity of a rotating matter. In a zero magnetic field, the electrical conductivity of a medium is isotropic in nature, which means that conductivity in all directions is the same. At a finite magnetic field, the isotropic conductivity breaks into three components, i.e., parallel, perpendicular, and Hall. Similarly, the isotropic conductivity can break into three similar parts during the transition from zero to finite rotation. The present article has demonstrated this picture by introducing its essential formalism part in Sec.~(\ref{sec:F}) and then by showing graphical descriptions in Sec.~(\ref{Results}) with final remarks in Sec.~(\ref{Summary}).
\section{Formalism}
\label{sec:F}
\begin{figure}
    \centering
    \includegraphics [scale=0.4]{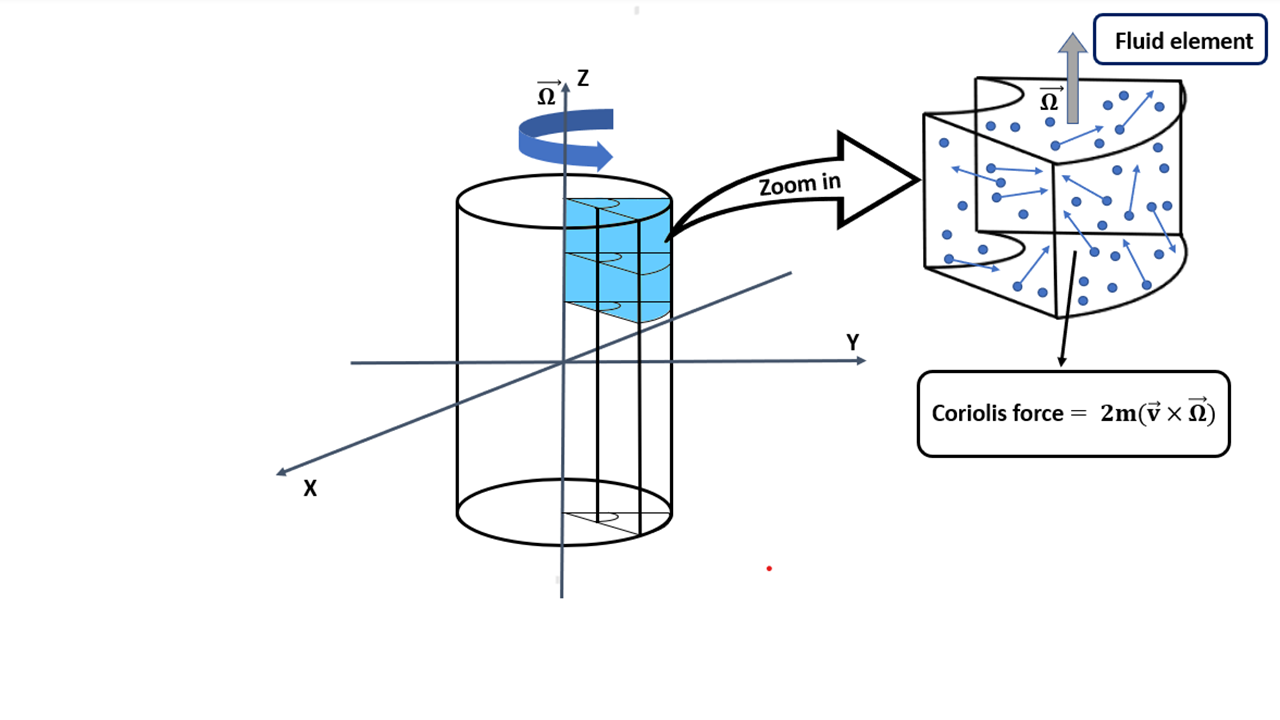}
    \caption{Schematic depiction of a rotating cylinder and Coriolis force on the rotating particles.}
    \label{fig:rot_cyl}
\end{figure}

We know the following operator equation linking the time derivative of an arbitrary vector from classical mechanics\cite{goldstein2011classical},
	\begin{equation}
	\left(\frac{d}{dt}\right)\equiv \left(\frac{d}{dt}\right)^{\prime}+\Vec{\Om}\times
	\label{B1}~,
	\end{equation} 
	where unprimed and primed time derivative operators mean the derivative with respect to time should be performed in space-fixed (inertial), and rotating frames, respectively. $\Vec{\Omega}$ is the angular velocity of the rotating frame with respect to the space-fixed frame. One can apply operators in Eq.~(\ref{B1}) two times on $\vec{r}$ 
 to get,
	\begin{equation}
	\Vec{a}= \Vec{a}^{\prime}+2(\Vec{\Om} \times \Vec{v}^{\prime})+ \Vec{\Om} \times (\Vec{\Om}\times\Vec{r}^{\prime})+ \Dot{\Vec{\Om}} \times \Vec{r}^{\prime}
    \label{B2}~,
	\end{equation}
 where the rotating frame quantities are identified with primes. We will work from a rotating frame, and our quantity of interest will be $\Vec{v}^{\prime}$. Since space fixed velocity $\Vec{v}$ will not appear anywhere in the subsequent calculation, for the sake of notational simplicity, we will drop the overhead prime and call $\Vec{v}^{\prime}$ as $\Vec{v}$ from now onwards. The first term in the RHS of Eq.(\ref{B2}) contributes to the Coriolis force. In Fig.~(\ref{fig:rot_cyl}), we have schematically displayed a fluid (with simple cylindrical geometry) rotating with angular velocity $\Vec{\Omega}$, which means the fluid has global vorticity $\Vec{\Omega}$. As is shown in the picture, fluid can be assumed to be built by many fluid elements. We have zoomed in on a particular element to illustrate that the particles inside it have a random part of velocity $\Vec{v}$ in addition to the velocity $\Vec{\Om}\times \Vec{r}$ provided by global vorticity $\Vec{\Omega}$. The Coriolis force faced by the fluid particles of mass $m$ is given by the equation $\Vec{F}=2m(\Vec{v}\times\Vec{\Om})$.  
 There is a similarity between a system under finite rotation and a system under the background of a finite magnetic field.
 The velocity-dependent Lorentz force $\Vec{F}=q(\Vec{v}\times\Vec{B})$, with $B_{i}\equiv Bb_{i}$ creates a momentum anisotropy in the medium which results in the anisotropic conductivity with three independent conductivity components. The same macroscopic structure of conductivity tensor can be expected for a medium under finite rotation or finite angular velocity $\Om_i\equiv\Om \om_i$ with the replacement of magnetic field unit vector $b_i$ by angular velocity unit vector $\om_i$.  

Now, let us begin with the microscopic expression of electrical current density given by kinetic theory~\cite{Dey:2020awu}: 
\begin{equation}
    j_{i} = qg\int{\frac{d^{3}\vec{p}}{{(2\pi})^{3}}} v_{i}~\delta{f}~,
    \label{A1}
\end{equation}
where $q$ is charge on the particles, $\delta{f}$ is small change from the equilibrium distribution function, $ j_{i}$ is electrical current density in the medium, and $g$ is the degeneracy associated with a particle. We know that the macroscopic expression of $j_i$ is basically Ohm's law $j_{i}= \sigma_{ij} \Tilde{E_{j}}$, where it can be connected with electric field $\Tilde{E_{j}}$
 via conductivity tensor $\sigma_{ij}$. Our aim will be to express $\delta {f}$ of Eq.~(\ref{A1}) in terms of $\Tilde{E_{j}}$, for which we will take help from the Boltzmann equation.

Boltzmann transport equation in the relaxation time approximation (RTA) can be written as~\cite{kremer2010int}:
\begin{equation}
    \frac{\partial f}{\partial t}+ \Vec{v}\cdot\frac{\partial f}{\partial \Vec{r}}+[q\Vec{\Tilde{E}}+2m(\Vec{v}\times\Vec{\Omega})]\cdot\frac{\partial f}{\partial\Vec{p}}= -\frac{\delta f}{\tau_c}
    \label{A5}~,
\end{equation}
where $\tau_{c}$ is the relaxation time, and $f=f^0+\delta f$.

We have ignored the contribution of the Centrifugal force in writing Eq.~(\ref{A5}) to maintain the equivalence between Lorentz force and Coriolis force in the main text. However, the calculation of electrical current in the presence of both Centrifugal and Coriolis forces can be done in principle by following the same methodology. We have added the details of the calculation in the presence of both forces in Appendix~(\ref{Appendix}).

The function $f^0$ is known as the local equilibrium distribution function, which for fermions is given by,
\begin{equation}
    f^0=\frac{1}{\exp{\frac{E-\mu(\Vec{r},t)-\Vec{u}(\Vec{r},t)\cdot\Vec{p}}{T(\Vec{r},t)}}+1}~,
    \label{A6}
\end{equation}
where, $\mu$ is the chemical potential, $E=\frac{1}{2}mv^{2}$ is the energy, $\Vec{u}(\Vec{r},t)$ is macroscopic velocity or the velocity of the quark fluid, $\Vec{p}=m\Vec{v}$ is particle momentum, and $\Vec{v}$ is particle velocity. One can obtain the following identities from the distribution~(\ref{A6}):
\begin{align}
    \frac{\partial f^0}{\partial E} &=\frac{-f^0(1-f^0)}{T}~,\nn\\
    \frac{\partial f^0}{\partial\Vec{p}}&=\frac{-f^0(1-f^0)(\Vec{v}-\Vec{u})}{T} =\frac{\partial f^0}{\partial E}(\Vec{v}-\Vec{u})~.\label{i1}
\end{align}
Since the present article is not structured for the calculation of viscosity or thermal conductivity, we will ignore the first two terms in the LHS Eq.~(\ref{A5}), which give rise to velocity and temperature gradient terms and rewrite Eq.~(\ref{A5}) as follows:
\begin{equation}
  [q\Vec{\Tilde{E}}+2m(\Vec{v}\times\Vec{\Omega})]\cdot\frac{\partial f}{\partial\Vec{p}}= -\frac{\delta f}{\tau_c}\label{i2}~.  
\end{equation}
Substituting the identities~(\ref{i1}) in Eq.~(\ref{i2}) and keeping the terms which are first order in $\delta f$, we have,
\begin{align*}
    q\Vec{\Tilde{E}}\cdot\frac{\partial f^0}{\partial\Vec{p}}+2m(\Vec{v}\times\Vec{\Omega})\cdot\frac{\partial f^0}{\partial \Vec{p}}+2m(\Vec{v}\times\Vec{\Omega})\cdot\frac{\partial\delta f}{\partial\Vec{p}} &=-\frac{\delta f}{\tau_{c}}\\
    \implies q\Vec{\Tilde{E}}\cdot\frac{\partial f^0}{\partial E}(\Vec{v}-\Vec{u})+2m(\Vec{v}\times\Vec{\Omega})\cdot\frac{\partial f^0}{\partial E}(\Vec{v}-\Vec{u})+2m(\Vec{v}\times\Vec{\Omega})\cdot\frac{\partial\delta f}{\partial\Vec{p}} &=-\frac{\delta f}{\tau_{c}}~.
\end{align*}
Since conductivity (in general, any transport coefficient) is independent of the fluid velocity, we will put $\Vec{u}=0$ in the above equation to get:
\begin{equation}
    \implies \frac{\partial f^0}{\partial E}\Vec{v}\cdot(q\Vec{\Tilde{E}})+2m(\Vec{v}\times\Vec{\Omega})\cdot\frac{\partial\delta f}{\partial\Vec{p}}=-\frac{\delta f}{\tau_{c}}~.
     \label{A7} 
\end{equation}
Let us assume that,~$\delta f=-\Vec{p}\cdot\vec{F} \big( \frac{\partial f^0}{\partial E} \big)$ with $\Vec{F}=\alpha \hat{e}+\beta \hat{\omega}+\gamma(\hat{e}\times\hat{\omega})$, where $\Vec{\Tilde{E}}=\Tilde{E}\hat{e},~ \Vec{\Omega}=\Omega\hat{\omega}$, and
$\alpha,\beta,\gamma$ are unknown constants, which would be determined from Eq.~(\ref{A7}). By substituting $\delta f$ in Eq.~(\ref{A7}), we get: 
\begin{align}
    &\frac{\partial f^0}{\partial E}\Vec{v}\cdot(q\Vec{\Tilde{E}})+2m\Vec{v}\cdot\left(\Vec{\Omega}\times\frac{\partial\delta f}{\partial\Vec{p}}\right)=\frac{-\delta f}{\tau_{c}}\nn\\   
   \implies&[q\Tilde{\Vec{E}}+2m(\Vec{F}\times\Vec{\Omega})]\cdot\Vec{v}\frac{\partial f^0}{\partial E}=\frac{m}{\tau_{c}} \Vec{F}\cdot\left(\Vec{v}\frac{\partial f^0}{\partial E}\right).
    \label{A8}
\end{align}
%
%
Substituting the result, $\Vec{F}\times\Vec{\Omega}=-\gamma\Omega\hat{e}+\gamma\Omega(\hat{\omega}\cdot\hat{e})\hat{\omega}+\alpha\Omega(\hat{e}\times\hat{\omega})$, in Eq.~(\ref{A8}) and using the fact that $\Vec{v}$ is arbitrary we have the identity,
\begin{eqnarray}
    \Big( \frac{q\Tilde{E}}{m}-2\Omega\gamma \Big) \hat{e}+2\gamma \Omega (\hat{\omega}\cdot\hat{e}) \hat{\omega} +2 \alpha \Omega (\hat{e} \times \hat{\omega}) =\frac{\alpha}{\tau_{c}}\hat{e}+\frac{\beta}{\tau_{c}}\hat{\omega}+\frac{\gamma}{\tau_{c}}(\hat{e}\times\hat{\omega})~.
    \label{A9}
\end{eqnarray}

By equating the coefficients of the linearly independent basis vectors, we get the following three equations: 
\begin{align}
 	\frac{q\tilde{E}}{m}-\frac{\gamma}{\tau_{\Omega}} &= \frac{\alpha}{\tau_c}~, &
 	\frac{\gamma}{\tau_{\Omega}}(\hat{\omega}.\hat{e}) &= \frac{\beta}{\tau_c}~, &
 	\frac{\alpha}{\tau_{\Omega}} &=\frac{\gamma}{\tau_c}~,\label{l3}
\end{align}
where we have quantified $\tau_{\Omega}\equiv \frac{1}{2\Omega}$.
Eq.~\ref{l3} can be simplified as:
\begin{align}
    \alpha &= \frac{\tau_c\big(\frac{q\tilde{E}}{m}\big)} {1+\big(\frac{\tau_c}{\tau_{\Omega}} \big)^2}~, & 
 	\gamma &=\frac{\tau_c \big(\frac{\tau_c}{\tau_{\Omega}}\big) \big(\frac{q\tilde{E}}{m} \big)} {1+\big(\frac{\tau_c}{\tau_{\Omega}}\big)^2}~, &
 	\beta &=\frac{\tau_c \big(\frac{\tau_c}{\tau_{\Omega}}\big)^2 (\hat{\omega}.\hat{e}) \big(\frac{q\tilde{E}}{m}\big)}{1+\big(\frac{\tau_c}{\tau_{\Omega}}\big)^2}~.\label{s1}
\end{align}
The Eqs.~(\ref{s1}) can be used to get the expression of $\delta f$ as:
\begin{align}
    \delta f &= -\vec{p}\cdot\vec{F} \frac{\partial f^0}{\partial E}\nn\\
	    &=-\frac{\partial f^0}{\partial E} m \vec{v}\cdot(\alpha\hat{e} +\beta\hat{\omega}+ \gamma(\hat{e}\times \hat{\omega}))\nn\\
	    &=-\frac{\partial f^0}{\partial E}\Bigg[\frac{q\tau_c\tilde{E}}{1+\big(\frac{\tau_c}{\tau_{\Omega}}\big)^2}\hat{e}\cdot\vec{v}+\frac{\tau_c \big(\frac{\tau_c}{\tau_{\Omega}}\big)^2 (\hat{\omega}\cdot\hat{e})q\tilde{E}}{1+\big(\frac{\tau_c}{\tau_{\Omega}}\big)^2} \hat{\omega}\cdot\vec{v}+\frac{\tau_c \big(\frac{\tau_c}{\tau_{\Omega}}\big) q \tilde{E}} {1+\big(\frac{\tau_c}{\tau_{\Omega}}\big)^2} (\hat{e} \times \hat{\omega})\cdot\vec{v}\Bigg]\nn\\
	    &= -\frac{\partial f^0}{\partial E} \bigg(\frac{q \tau_c}{1+ \big( \frac{\tau_c} {\tau_{\Omega}} \big)^2} \bigg) \bigg[\delta_{jl}+ \Big( \frac{\tau_c}{\tau_{\Omega}} \Big)^2 \omega_j\omega_l+\Big(\frac{\tau_c}{\tau_{\Omega}}\Big) \ep_{ljk} \omega_k \bigg] \tilde{E}_j v_l~. \label{C0}
\end{align}
Now substituting expression of $\delta f$ obtained in Eq.(\ref{C0}) in Eq.(\ref{A1}) current density $j_i$ is given by,
\begin{equation}
 	j_{i}=-qg \int \frac{d^3\Vec{p}}{(2\pi)^3}v_{i}\frac{\partial f^0}{\partial E}\bigg(\frac{q\tau_c}{1+\Big(\frac{\tau_c}{\tau_{\Omega}}\Big)^2}\bigg)\bigg[\delta_{jl}+\Big(\frac{\tau_c}{\tau_{\Omega}}\Big)^2\omega_j\omega_l+\Big(\frac{\tau_c}{\tau_{\Omega}}\Big) \ep_{ljk}\omega_k\bigg]\tilde{E}_jv_{l}~.
 	\label{A10}
\end{equation}
We can calculate the angular average,
\begin{equation}
    \int d^3\Vec{p}~v_iv_l=\int 4\pi p^2dp\left(\frac{v^2}{3}\right)\delta_{il}=\int d^{3}p\left(\frac{v^2}{3}\right)\delta_{il}~, 
    \label{A11}
\end{equation} 	
where $d^{3}p\equiv 4\pi p^2~dp$.
Substituting the result of Eq.~(\ref{A11}) in Eq.~(\ref{A10}) we have:
\begin{align}
  j_{i}&=-q^2g\int \frac{d^{3}p}{(2\pi)^3} \frac{\partial f^0}{\partial E} \bigg(\frac{\tau_c}{1+\Big(\frac{\tau_c}{\tau_{\Omega}}\Big)^2}\bigg)\frac{v^2}{3}\delta_{il}\bigg[\delta_{jl}+\Big(\frac{\tau_c}{\tau_{\Omega}}\Big)^2\omega_j\omega_l+\Big(\frac{\tau_c}{\tau_{\Omega}}\Big)\ep_{ljk}\omega_k \bigg]\tilde{E}_j\nn\\
      &=-q^2g \int \frac{d^{3}p}{(2\pi)^3}\frac{\partial f^0}{\partial E}\bigg(\frac{\tau_c}{1+\Big(\frac{\tau_c}{\tau_{\Omega}}\Big)^2}\bigg) \frac{v^2}{3} \bigg[\delta_{ij}+\Big(\frac{\tau_c}{\tau_{\Omega}}\Big)^2\omega_j\omega_i+\Big( \frac{\tau_c}{\tau_{\Omega}} \Big) \ep_{ijk} \omega_k \bigg] \tilde{E}_j~.
      \label{C1}
\end{align}
 By comparing the RHS of Eq.(\ref{C1}) with the expression $j_i=\sigma_{ij}\tilde{E}_j$ we obtain the conductivity of the system $\sigma_{ij}$ as:
\begin{equation}
  \sigma_{ij} = -g q^2 \int \frac{d^{3}p}{(2\pi)^3}\frac{\partial f^0}{\partial E} \bigg(\frac{\tau_c}{1+\big(\frac{\tau_c}{\tau_{\Omega}}\big)^2}\bigg)\frac{v^2}{3} \bigg[\delta_{ij}+\Big(\frac{\tau_c}{\tau_{\Omega}}\Big)^2\omega_j\omega_i+\Big( \frac{\tau_c}{\tau_{\Omega}}\Big)\ep_{ijk} \omega_k \bigg]\label{gct}~.  
\end{equation}
 We can re-express Eq.~(\ref{gct}) as follows:	
\begin{equation}
 \sigma_{ij} = \sigma_0 \delta_{ij} +\sigma_1\ep_{ijk}\omega_k +\sigma_2\omega_i\omega_j~,
\text{with}~\sigma_n = \frac{g q^2}{T}\int \frac{d^{3}p}{(2\pi)^3}\frac{\tau_c\big(\frac{\tau_c}{\tau_{\Omega}}\big)^n} {1+\big(\frac{\tau_c}{\tau_{\Omega}}\big)^2}\times \frac{v^2}{3}f^0(1-f^0)\label{con_T}~.    
\end{equation}
$\sigma_0, \sigma_1, \sigma_2$ are scalars that make up the conductivity tensor.
The current density can also be written as, 
\begin{equation}
      \Vec{j}=\sigma_{0}\Vec{\tilde{E}}+\sigma_{1}(\Vec{\tilde{E}}\times \hat{\omega})+\sigma_{2}(\hat{\omega}\cdot\Vec{\tilde{E}})\hat{\omega}~.
      \label{A14}
\end{equation}
For angular velocity in the z-direction, the conductivity matrix can be written as:
\begin{equation}
  [\sigma]=\begin{pmatrix}
      \sigma_{0} & \sigma_{1} & 0\\
      -\sigma_{1} & \sigma_{0} & 0\\
      0 & 0 & \sigma_{0}+\sigma_{2}
  \end{pmatrix}~.
  \label{A15}
\end{equation}
The conductivity integral (\ref{con_T}) with the substitution  $E=\frac{1}{2}mv^{2}$ becomes:
\begin{eqnarray}
    \sigma_{n}&=&\frac{2\sqrt{m}gq^{2}}{3\pi^{2}T\sqrt{2}}\frac{\tau_c\left(\frac{\tau_c}{\tau_{\Omega}}\right)^n}{1+\left(\frac{\tau_c}{\tau_{\Omega}}\right)^2}\int
    {E^{3/2} f^0(1-f^0)} {dE}\nn\\
  \implies \sigma_{n}&=&\frac{2\sqrt{m}gq^{2}}{3\pi^{2}T\sqrt{2}}\frac{\tau_c\left(\frac{\tau_c}{\tau_{\Omega}}\right)^n}{1+\left(\frac{\tau_c}{\tau_{\Omega}}\right)^2}\int
   T\frac{\partial f^{0}}{\partial \mu} {E^{3/2}} {dE}\nn\\
   \implies \sigma_{n}&=&\frac{2\sqrt{m}gq^{2}}{3\pi^{2}\sqrt{2}}\frac{\tau_c\left(\frac{\tau_c}{\tau_{\Omega}}\right)^n}{1+\left(\frac{\tau_c}{\tau_{\Omega}}\right)^2}\frac{\partial}{\partial\mu}\int\frac{E^{3/2}}{e^{(E-\mu)/T}+1} {dE}\nn\\
    \implies \sigma_{n}&=&\frac{\sqrt{m\pi}gq^{2}}{2\pi^{2}\sqrt{2}}\frac{\tau_c\left(\frac{\tau_c}{\tau_{\Omega}}\right)^n}{1+\left(\frac{\tau_c}{\tau_{\Omega}}\right)^2}T^{\frac{5}{2}}\frac{\partial }{\partial\mu}f_{5/2}(A)\nn\\
     \implies \sigma_{n}&=&\frac{\sqrt{m\pi}gq^{2}}{2\pi^{2}\sqrt{2}}\frac{\tau_c\left(\frac{\tau_c}{\tau_{\Omega}}\right)^n}{1+\left(\frac{\tau_c}{\tau_{\Omega}}\right)^2}T^{\frac{3}{2}}f_{3/2}(A)~,\label{gen_cn}
\end{eqnarray}
 where we have used the results that $\frac{\partial f^{0}}{\partial \mu}=f^0(1-f^0)\frac{1}{T}$, $f_j(A)=\frac{1}{\Gamma(j)}\int_{0}^{\infty}\frac{x^{j-1}dx}{A^{-1}e^{x}+1}$, and  $\frac{\partial f_{5/2}(e^{\mu/T})}{\partial(\mu/T)}=f_{3/2}(e^{\mu/T})=f_{3/2}(A) \text{ with }A=e^{\mu/T}$.  
Comparing the result with the conductivity tensor in the presence of finite magnetic field \cite{Dey:2021fbo,Dey:2020awu,Dey:2019vkn,Dey:2019axu,Ghosh:2019ubc} and following the similarities, one can define $\sigma_{0}\equiv\sigma_{\perp}$, $\sigma_{0}+\sigma_{2}\equiv\sigma_{||}$, and $\sigma_{1}\equiv\sigma_{\times}$. $\sigma_{||}$ is also equal to $\sigma$, which is the conductivity in the absence of vorticity~\cite{Dey:2019axu}. Hence, we have the following:
\begin{align}
     \sigma_{||}&=\frac{g\sqrt{m}q^2}{(2\pi)^{3/2}} ~ \tau_c ~ T^{\frac{3}{2}}f_{3/2}(A)~,\nn\\
    \sigma_{\perp}&=\frac{g\sqrt{m}q^2}{(2\pi)^{3/2}}\frac{\tau_c}{1+\big(\frac{\tau_c}{\tau_{\Omega}}\big)^2}T^{\frac{3}{2}}f_{3/2}(A)~,\nn\\
    \sigma_{\times}&=\frac{g\sqrt{m}q^2}{(2\pi)^{3/2}}\frac{\tau_c\big(\frac{\tau_c}{\tau_{\Omega}} \big)}{1+\big(\frac{\tau_c}{\tau_{\Omega}}\big)^2}T^{\frac{3}{2}}f_{3/2}(A)~.\label{gen_cp}
\end{align}
\section{Results}
\label{Results}
Our general expressions~(\ref{gen_cp}) are also valid for degenerate matters found in systems like white dwarfs by taking limit $T\xrightarrow{}0$,
\begin{align}
    \sigma_{||}&=\frac{4g}{3\sqrt{\pi}}\frac{\sqrt{m}q^2}{(2\pi)^{3/2}} ~ \tau_c ~ \mu^{\frac{3}{2}}~,\nn\\
    \sigma_{\perp}&=\frac{4g}{3\sqrt{\pi}}\frac{\sqrt{m}q^2}{(2\pi)^{3/2}}\frac{\tau_c}{1+\big(\frac{\tau_c}{\tau_{\Omega}}\big)^2}\mu^{\frac{3}{2}}~,\nn\\
    \sigma_{\times}&=\frac{4g}{3\sqrt{\pi}}\frac{\sqrt{m}q^2}{(2\pi)^{3/2}}\frac{\tau_c\big(\frac{\tau_c}{\tau_{\Omega}} \big)}{1+\big(\frac{\tau_c}{\tau_{\Omega}}\big)^2}\mu^{\frac{3}{2}}~.\label{gen_cph} 
\end{align}
We got the general expressions ~(\ref{gen_cp}) for the component conductivities in the previous section, which are valid for any chemical potential ($\mu$), temperature ($T$), and angular velocity ($\Om$). The expressions obtained here can be promptly applied to non-relativistic fluids, namely, the condensed matter systems where the quantities $\mu$, $T$, and $\Om$ are of the order of eV in the natural unit. However, our aimed systems belong to relativistic high energy systems, namely, quark-gluon plasma (QGP) and compact stars where the quantities $\mu$, $T$, and $\Om$ are of the order of MeV. One can identify two extreme domains from the $T-\mu$ diagram of the quark-hadron phase transition, specifically, the net zero baryon density picture (at $\mu\xrightarrow{}0$) occurs in the early universe, which is reproducible in LHC and RHIC experiments, and the degenerate nucleon or quark matter picture that occurs in white dwarfs and neutron stars (at $T\xrightarrow{}0$). Our general expressions~(\ref{gen_cp}) can be applied to both pictures by putting $\mu=0$ or $T=0$. Although in such cases, there would be some ubiquitous overestimation due to using a non-relativistic framework for the relativistic systems. Our immediate future objective is to develop a relativistic framework that can be applied safely for the above cases.\\
\begin{figure}
    \centering
    \includegraphics[scale=0.30]{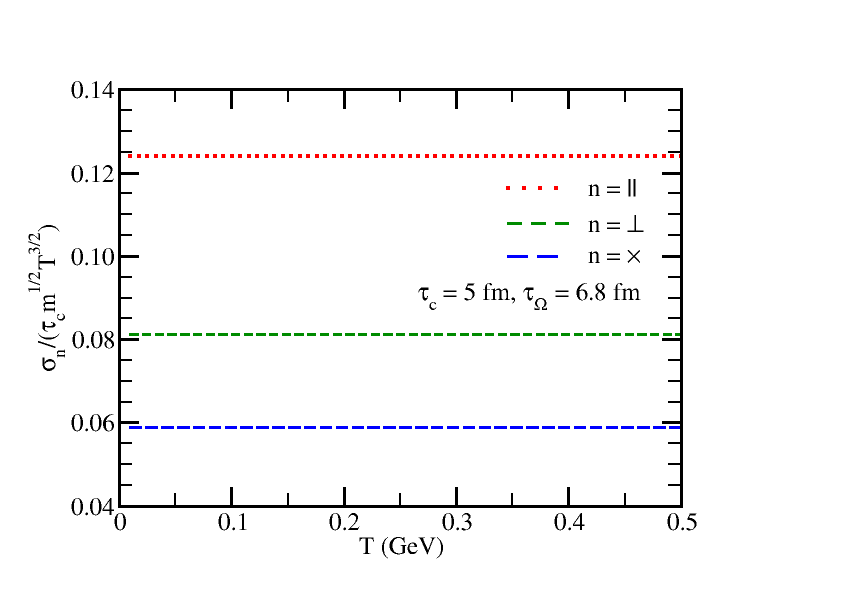}
    \caption{Normalized electrical conductivity components vs temperature.}
    \label{fig:Temp}
\end{figure}
\begin{figure}
    \centering
   \includegraphics[scale=0.30]{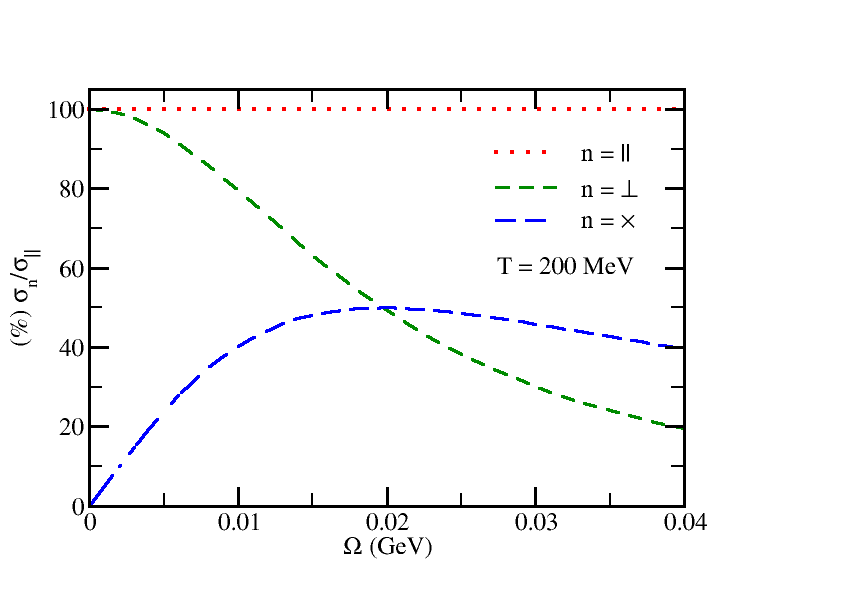}
    \caption{Variation of (normalized) conductivity components with respect to $\Omega$.}
    \label{fig:Omega}
\end{figure}
\begin{figure}
    \centering
    \includegraphics[scale= 0.22]{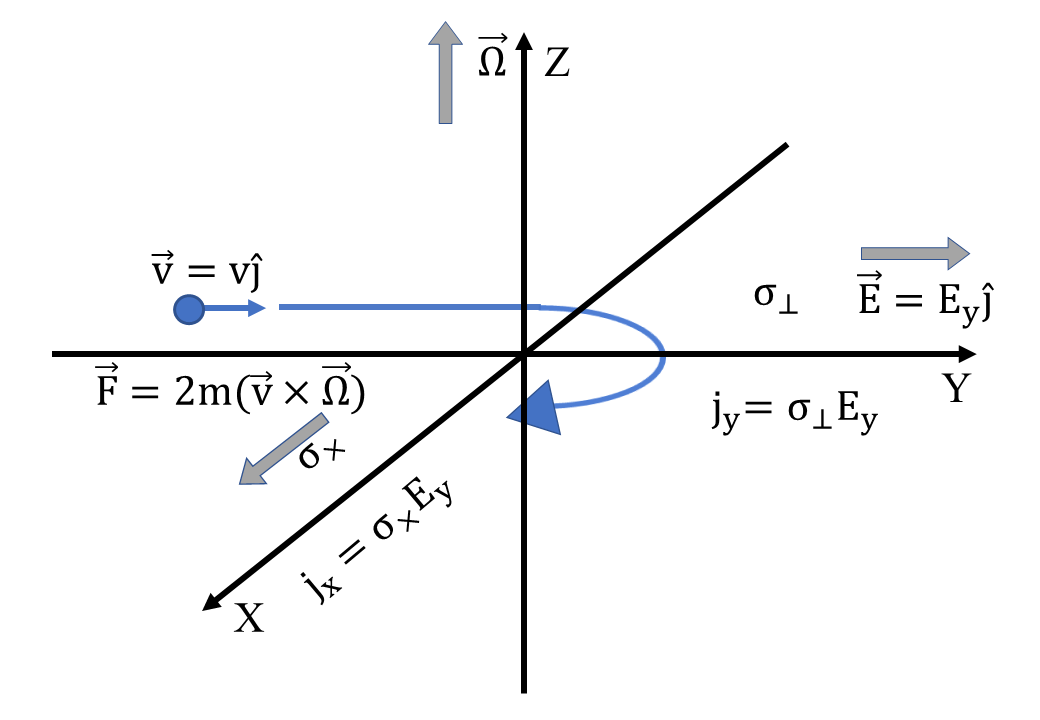}
    \includegraphics[scale= 0.22]{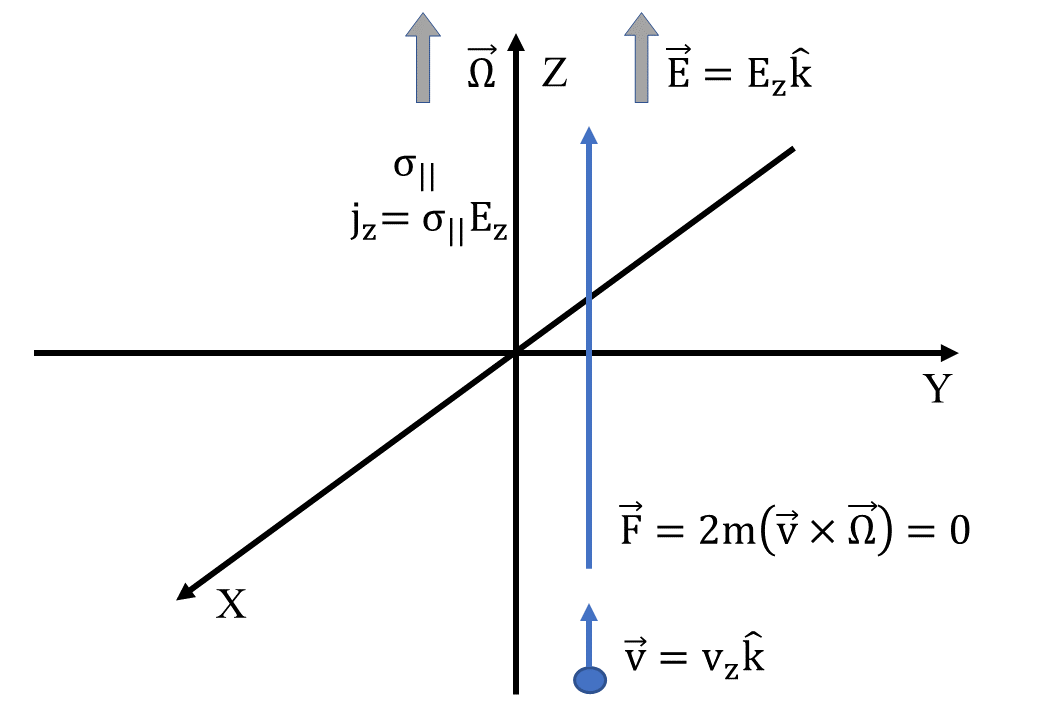}
    \caption{Representation of different components of conductivity.}
    \label{fig:an_con}
\end{figure}

By taking the limit $\mu\xrightarrow{} 0$ in Eq.~(\ref{gen_cp}) we get:
\begin{align}
    \sigma_{||}&=0.29g~\frac{\sqrt{m}q^2}{(2\pi)^{3/2}} ~ \tau_c ~ T^{\frac{3}{2}}\zeta(3/2)~,\nn\\
    \sigma_{\perp}&=0.29g~\frac{\sqrt{m}q^2}{(2\pi)^{3/2}}\frac{\tau_c}{1+\big(\frac{\tau_c}{\tau_{\Omega}}\big)^2}T^{\frac{3}{2}}\zeta(3/2)~,\nn\\
    \sigma_{\times}&=0.29g~\frac{\sqrt{m}q^2}{(2\pi)^{3/2}}\frac{\tau_c\big(\frac{\tau_c}{\tau_{\Omega}} \big)}{1+\big(\frac{\tau_c}{\tau_{\Omega}}\big)^2}T^{\frac{3}{2}}\zeta(3/2)~,\label{spcl_c1}
\end{align}
since $f_{3/2}(1)=(1-\frac{1}{2^{1/2}})\zeta(3/2).$ 
We have used the results of Eq.~(\ref{spcl_c1}) to plot $\sigma_{||,\perp,\times}/\sqrt{m}T^{3/2}\tau_c$ with respect to $T$ in Fig.~(\ref{fig:Temp}). In  Fig.~(\ref{fig:Temp}), we get all horizontal lines because all the component conductivities are proportional to $T^{3/2}$. We have used $m=0.002$ GeV by assuming the medium constituent particle is u quark (but non-relativistic in nature). Typical quantum chromodynamics (QCD) interaction time scale is of the order of fm, so we take the relaxation time scale as $\tau_c=5$ fm.
Regarding the value of $\tau_\Om=1/(2\Om)$, we need the knowledge of the order of magnitude for average vorticity, produced in peripheral heavy ion collisions. Different model calculations~\cite{PhysRevC.93.064907, PhysRevC.94.044910,PhysRevC.95.054915,Huang:2020dtn} have predicted the average kinematic vorticity of the quark-gluon plasma, which depends on impact parameters, beam energy, and time. 
For example, Ref.~(\cite{PhysRevC.94.044910}) has indicated that peak value of ~$\Omega$~ can have a range from $0.02$ to $0.14 \text{ fm}^{-1}$ or from $0.003$ GeV to $0.027$ GeV~. This $\Omega$ range corresponds to a $\tau_{\Omega}$ range of $3.57$ fm~to $25$ fm~. We have considered a value $\tau_\Om=6.8$ fm, which correspond to angular velocity $\Om=\frac{1}{2\tau_\Om}=0.014$ GeV. 


The conductivity component $\sigma_{||}$ is proportional to relaxation time, i.e., $\sigma_{||}\propto \tau_c$. Other components are proportional to their effective relaxation times, i.e.,   $\sigma_{\perp} \propto \tau_{\perp}$, and $\sigma_{\times}\propto \tau_{\times}$, where the effective relaxation times are defined as:
\begin{align*}
  \tau_{\perp}&=\frac{\tau_c}{1+\big(\frac{\tau_c}{\tau_{\Omega}}\big)^2}~,\\
  \tau_{\times}&=\frac{\tau_c\big(\frac{\tau_c}{\tau_{\Omega}} \big)}{1+\big(\frac{\tau_c}{\tau_{\Omega}}\big)^2}~.
\end{align*}

One can easily understand that the non-zero ratio of $\tau_c$ and $\tau_{\Om}$ make different relaxation times unequal. The magnitude of $\tau_c/\tau_{\Om}$ determines the ordering of $\sigma_{\perp}$ and $\sigma_{\times}$. To see the relative variation of different components of conductivities with vorticity, we plotted the percentile of normalized conductivities($\sigma_{{||},{\perp},{\times}}/\sigma$) in Fig.~(\ref{fig:Omega}) with respect to $\Omega$ at zero chemical potential with $\tau_c =5$ fm and $T=200$ MeV. $\sigma_{||}$ does not change with change in $\Om$. But it can be seen in the figure that $\sigma_\perp$ decreases with an increase in $\Omega$ in the whole range of $\Omega$. On the other hand, $\sigma_\times$ initially increases but then decreases with $\Omega$. It should be noticed that for low-value vorticity, $\sigma_\perp$ is more dominant than $\sigma_\times$, whereas, for the high vorticity value, $\sigma_\times$ is more dominant than  $\sigma_\perp$ with the crossover happening around $\Omega=0.02$ GeV. One can see from Fig.~(\ref{fig:Omega}) that $\sigma_{\perp}$ merge to $\sigma_{||}$ or $\sigma$ in the absence of vorticity, i.e., $\sigma_{\perp}(\Om\xrightarrow{} 0)=\sigma$ whereas $\sigma_{\times}$ vanishes as one goes to the limit of zero vorticity, i.e., $\sigma_{\times}(\Om\xrightarrow{} 0)=0$. It is clear from the above fact that finite angular velocity or global vorticity creates anisotropy in the electrical conductivity of a system like a magnetic field.

We have schematically depicted different components of conductivity tensor in Fig.~(\ref{fig:an_con}). The scenario is similar to the scenario in Refs.~(\cite{Dey:2021fbo,Dey:2020awu,Dey:2019vkn,Dey:2019axu,Ghosh:2019ubc}) if one replaces the magnetic field with vorticity. In the left panel of the figure, the direction of the electric field and vorticity or angular velocity have been chosen to be along the positive  Y and Z-axis, respectively. Here, along with the conduction along the positive Y-axis with conductivity $\sigma_{\perp}$, a flow of charge along the X-direction also occurs due to the action of Coriolis force. The current density in the X-direction is proportional to the $\sigma_\times$, i.e., $j_{x}=\sigma_{\times}E_{y}$. In the right panel of the figure, the direction of the electric field and vorticity or angular velocity have been chosen to be in the Z-direction. Since the conduction along this direction is unaffected by the vorticity due to the null contribution of the Coriolis force, we obtain current density in the Z-direction with the isotropic conductivity $\sigma$, i.e., $j_{z}=\sigma E_{z}$.
 Similar to the case of the quantum hall effect in the quantum picture of the magnetic field, the quantum hall effect in the presence of rotation may be seen, and the corresponding conductivity may be derived by solving the Schrodinger equation in a rotating frame and using the energy eigenvalues in the distribution function. One of our current plans is to calculate the transport coefficients of a medium, including the energy quantization of the particles, by considering the quantum aspects of potentials present in a rotating reference frame. 
 
\section{Summary}
\label{Summary}
In summary, we have explored the equivalence nature of the Lorentz force with the Coriolis force to calculate the electrical conductivity of a medium with finite global vorticity. In the absence of a magnetic field, the system only has an isotropic conductivity $\sigma$, which breaks into parallel, perpendicular, and Hall components in the presence of a magnetic field due to the action of Lorentz force. Similarly, the conductivity of a medium turns from isotropic to anisotropic during the transition from zero to nonzero global vorticity due to the action of the Coriolis force. Similar to the magnetic field case, the conductivity in the parallel direction remains unchanged, whereas, in the perpendicular and Hall directions, one obtains different conductivities proportional to two different effective relaxation times. The effective relaxation times are made up of relaxation time and cyclotron-type time periods. The mathematical steps in calculating conductivity in the presence of global vorticity are quite similar to those of the magnetic field case. Here, we only considered calculating the electrical conductivity in the presence of vorticity; nevertheless, its equivalence with the magnetic field case has been stated throughout. Firstly, the microscopic quantity- the deviation from the equilibrium distribution, is guessed with three unknown constants. Secondly, the deviation is found by substituting it in the relaxation time approximated Boltzmann equation with the Coriolis force term. Then, it is used to obtain the microscopic expression of current density and is compared with the macroscopic definition of current density to get the conductivity tensor.  
\section{Acknowledgement}
AD gratefully acknowledges the MoE, Govt. of India. CWA acknowledges the DIA programme. This work was partly supported by the Doctoral Fellowship in India (DIA) programme of the Ministry of Education, Government of India. JD gratefully acknowledges the DAE-DST, Govt. of India funding under the mega-science project – “Indian participation in the ALICE experiment at CERN” bearing Project No. SR/MF/PS-02/2021- IITI (E-37123). AD thanks Anita Tamang for her help connected with this work.  SG  thanks  Arghya Mukherjee and Deeptak Biswas for the valuable discussion during the initial stage of the article.
\section{Appendix}
\label{Appendix}
\textbf{Inclusion of Centrifugal Force:} To see the effect of Centrifugal Force on the current, one may add the centrifugal force as a force term in the LHS of Eq.~(\ref{i2}) as follows:
\bea
 &&[q\Vec{\Tilde{E}}+2m(\Vec{v}\times\Vec{\Omega})-m\Vec{\Omega}\times(\Vec{\Omega}\times\Vec{r})]\cdot\frac{\partial f}{\partial\Vec{p}}= -\frac{\delta f}{\tau_c}~\nn\\
\implies&&[q\Vec{\Tilde{E}}-m\Vec{\Omega}\times(\Vec{\Omega}\times\Vec{r})]\cdot\frac{\partial f^0}{\partial\Vec{p}}+2m(\Vec{v}\times\Vec{\Omega})\cdot\frac{\partial f^{0}+\delta f}{\partial\Vec{p}}= -\frac{\delta f}{\tau_c}~\nn\\
\implies && [q\Vec{\Tilde{E}}-m((\Vec{\Omega}\cdot\Vec{r})\Vec{\Omega}-\Omega^{2}\Vec{r})]\cdot\frac{\partial f^0}{\partial\Vec{p}}+2m(\Vec{v}\times\Vec{\Omega})\cdot\frac{\partial f^{0}+\delta f}{\partial\Vec{p}}= -\frac{\delta f}{\tau_c}~\nn\\
\implies && [q\Vec{\Tilde{E}}-m((\Vec{\Omega}\cdot\Vec{r})\Vec{\Omega}-\Omega^{2}\Vec{r})]\cdot\frac{\partial f^0}{\partial E}(\Vec{v}-\Vec{u})+2m(\Vec{v}\times\Vec{\Omega})\cdot\frac{\partial f^0}{\partial E}(\Vec{v}-\Vec{u})+2m(\Vec{v}\times\Vec{\Omega})\cdot\frac{\partial\delta f}{\partial\Vec{p}}= -\frac{\delta f}{\tau_c}~.\label{ap1}
 \eea
 Note that, in heavy ion collisions, the angular velocity created is comparatively weaker than the produced magnetic field; therefore, ignoring the higher order correction in the $\delta f$ due to centrifugal force is reasonable. 
 Since conductivity (in general, any transport coefficient) is independent of the fluid velocity, we will put $\Vec{u}=0$ in the Eq.~(\ref{ap1}) to get:
 \begin{align}
 && [q\Vec{\Tilde{E}}-m((\Vec{\Omega}\cdot\Vec{r})\Vec{\Omega}-\Omega^{2}\Vec{r})]\cdot\frac{\partial f^0}{\partial E}\Vec{v}+2m(\Vec{v}\times\Vec{\Omega})\cdot\frac{\partial f^0}{\partial E}\Vec{v}+2m(\Vec{v}\times\Vec{\Omega})\cdot\frac{\partial\delta f}{\partial\Vec{p}}= -\frac{\delta f}{\tau_c}~,\nn\\
&&[q\Vec{\Tilde{E}}-m((\Vec{\Omega}\cdot\Vec{r})\Vec{\Omega}-\Omega^{2}\Vec{r})]\cdot\frac{\partial f^0}{\partial E}\Vec{v}+2m(\Vec{v}\times\Vec{\Omega})\cdot\frac{\partial\delta f}{\partial\Vec{p}}= -\frac{\delta f}{\tau_c}~.\label{ap2}
 \end{align}
 Let us assume that,~$\delta f=-\Vec{p}\cdot\vec{F} \big( \frac{\partial f^0}{\partial E} \big)$ with $\Vec{F}=\alpha \hat{e}+\beta \hat{\omega}+\gamma(\hat{e}\times\hat{\omega})$, where $\Vec{\Tilde{E}}=\Tilde{E}\hat{e},~ \Vec{\Omega}=\Omega\hat{\omega}$, and
$\alpha,\beta,\gamma$ are unknown constants.
The Eq.(\ref{ap2}) with the substitution of $\delta f$ becomes:
\begin{align}
    [\frac{q\Vec{\Tilde{E}}}{m}-((\Vec{\Omega}\cdot\Vec{r})\Vec{\Omega}-\Omega^{2}\Vec{r})+2(\Vec{F}\times\Vec{\Omega})]\cdot\frac{\partial f^0}{\partial E}\Vec{v}= \frac{1}{\tau_{c}} \Vec{F}\cdot\left(\Vec{v}\frac{\partial f^0}{\partial E}\right)~.\label{ap3}
\end{align}
We have three linearly independent vectors, i.e, $\hat{e}$, $\hat{\omega}$, and $\hat{\omega}\times\hat{e}$~. One may, without loss of generality, assume that $\hat{\omega}=\hat{k},\frac{\hat{e}\times\hat{\omega}}{|\hat{e}\times\hat{\omega}|}= \hat{j} $, and  $\frac{(\hat{e}\times\hat{\omega})\times\hat{\omega}}{|\hat{e}\times\hat{\omega}|}=\frac{\hat{\omega}\times(\hat{\omega}\times\hat{e})}{|\hat{e}\times\hat{\omega}|}=\hat{i}$, where $\hat{i}, \hat{j}$, and $\hat{k}$ are the unit vectors along X, Y, and Z axes, respectively. Now we will express the expression of centrifugal force in terms of the linear combination of  $\hat{e}$, $\hat{\omega}$, and $\hat{e}\times\hat{\omega}$~.\\
\begin{eqnarray}
  \text{Centrifugal acceleration}&=& \Omega^{2}(\Vec{r}-(\hat{\omega}\cdot\Vec{r})\hat{\omega})\nn\\
 &=& \Omega^{2}(x\hat{i}+y\hat{j})\nn\\ 
 &=&\Omega^{2}\left[x \frac{\hat{\omega}\times(\hat{\omega}\times\hat{e})}{|\hat{e}\times\hat{\omega}|} +y\frac{\hat{e}\times\hat{\omega}}{|\hat{e}\times\hat{\omega}|}\right]\nn\\
 &=&\frac{1}{4\tau_{\Om}^{2}}\left[-\frac{x}{|\hat{e}\times\hat{\omega}|}\hat{e}+\frac{x(\hat{\omega}\cdot\hat{e})}{|\hat{e}\times\hat{\omega}|}\hat{\omega}+\frac{y}{|\hat{e}\times\hat{\omega}|}\hat{e}\times\hat{\omega}\right]~. \label{ap4}
 \end{eqnarray}  
 Further one notices $\Vec{F}\times\Vec{\Omega}=-\gamma\Omega\hat{e}+\gamma\Omega(\hat{\omega}\cdot\hat{e})\hat{\omega}+\alpha\Omega(\hat{e}\times\hat{\omega})$.
 By equating the coefficients of the linearly independent basis vectors from both sides of Eq.~(\ref{ap3}), we get the following three equations: 
\begin{align}
 	\frac{q\tilde{E}}{m}-\frac{\gamma}{\tau_{\Omega}}-\frac{x}{4\tau_{\Omega}^{2}|\hat{e}\times\hat{\omega}|} &= \frac{\alpha}{\tau_c}~, &
 	\frac{\gamma}{\tau_{\Omega}}(\hat{\omega}.\hat{e})+\frac{x(\hat{\omega}\cdot\hat{e})}{4\tau_{\Omega}^{2}|\hat{e}\times\hat{\omega}|} &= \frac{\beta}{\tau_c}~, &
 	\frac{\alpha}{\tau_{\Omega}}+\frac{y}{4\tau_{\Omega}^{2}|\hat{e}\times\hat{\omega}|} &=\frac{\gamma}{\tau_c}~.\label{ap5}
\end{align}
The above equations can be simplified to give:
\begin{align}
    \alpha &= \frac{\tau_c\big(\frac{q\tilde{E}}{m}\big)} {1+\big(\frac{\tau_c}{\tau_{\Omega}} \big)^2}-\frac{\frac{\tau_c}{\tau_{\Omega}}}{4\tau_{\Om}|\hat{e}\times\hat{\om}|\big(1+\big(\frac{\tau_c}{\tau_{\Omega}} \big)^2\big)}(x+\frac{\tau_c}{\tau_{\Om}}y)~,\nn\\
 	\gamma &=\frac{\tau_c \big(\frac{\tau_c}{\tau_{\Omega}}\big) \big(\frac{q\tilde{E}}{m} \big)} {1+\big(\frac{\tau_c}{\tau_{\Omega}}\big)^2}+\frac{\frac{\tau_c}{\tau_{\Omega}}}{4\tau_{\Om}|\hat{e}\times\hat{\om}|\big(1+\big(\frac{\tau_c}{\tau_{\Omega}} \big)^2\big)}(-x\frac{\tau_c}{\tau_{\Om}}+y)~,\nn\\
 	\beta &=\frac{\tau_c \big(\frac{\tau_c}{\tau_{\Omega}}\big)^2 (\hat{\omega}.\hat{e}) \big(\frac{q\tilde{E}}{m}\big)}{1+\big(\frac{\tau_c}{\tau_{\Omega}}\big)^2}+\frac{\frac{\tau_c}{\tau_{\Omega}}\hat{\omega}.\hat{e}}{4\tau_{\Om}|\hat{e}\times\hat{\om}|\big(1+\big(\frac{\tau_c}{\tau_{\Omega}} \big)^2\big)}(x+\frac{\tau_c}{\tau_{\Om}}y)~.\label{ap6}
\end{align}
\begin{eqnarray}
 \delta f &=& -\vec{p}\cdot\vec{F} \frac{\partial f^0}{\partial E}\nn\\
	    &=&-\frac{\partial f^0}{\partial E} m \vec{v}\cdot(\alpha\hat{e} +\beta\hat{\omega}+ \gamma(\hat{e}\times \hat{\omega}))\nn\\ 
            &=&-\frac{\partial f^0}{\partial E} m \vec{v}\cdot\bigg[\big[\frac{\tau_c\big(\frac{q\tilde{E}}{m}\big)} {1+\big(\frac{\tau_c}{\tau_{\Omega}} \big)^2}-\frac{\frac{\tau_c}{\tau_{\Omega}}}{4\tau_{\Om}|\hat{e}\times\hat{\om}|\big(1+\big(\frac{\tau_c}{\tau_{\Omega}} \big)^2\big)}(x+\frac{\tau_c}{\tau_{\Om}}y)\big]\hat{e}+\big[\frac{\tau_c \big(\frac{\tau_c}{\tau_{\Omega}}\big)^2 (\hat{\omega}.\hat{e}) \big(\frac{q\tilde{E}}{m}\big)}{1+\big(\frac{\tau_c}{\tau_{\Omega}}\big)^2}+\frac{\frac{\tau_c}{\tau_{\Omega}}\hat{\omega}.\hat{e}}{4\tau_{\Om}|\hat{e}\times\hat{\om}|\big(1+\big(\frac{\tau_c}{\tau_{\Omega}} \big)^2\big)}(x+\frac{\tau_c}{\tau_{\Om}}y)\big]\hat{\omega}\nn\\
             &&+\big[\frac{\tau_c \big(\frac{\tau_c}{\tau_{\Omega}}\big) \big(\frac{q\tilde{E}}{m} \big)} {1+\big(\frac{\tau_c}{\tau_{\Omega}}\big)^2}+\frac{\frac{\tau_c}{\tau_{\Omega}}}{4\tau_{\Om}|\hat{e}\times\hat{\om}|\big(1+\big(\frac{\tau_c}{\tau_{\Omega}} \big)^2\big)}(-x\frac{\tau_c}{\tau_{\Om}}+y)](\hat{e}\times \hat{\omega}))\bigg]\nn\\
              &=&-\frac{\partial f^0}{\partial E}\Bigg[\frac{q\tau_c\tilde{E}}{1+\big(\frac{\tau_c}{\tau_{\Omega}}\big)^2}\hat{e}\cdot\vec{v}+\frac{\tau_c \big(\frac{\tau_c}{\tau_{\Omega}}\big)^2 (\hat{\omega}\cdot\hat{e})q\tilde{E}}{1+\big(\frac{\tau_c}{\tau_{\Omega}}\big)^2} \hat{\omega}\cdot\vec{v}+\frac{\tau_c \big(\frac{\tau_c}{\tau_{\Omega}}\big) q \tilde{E}} {1+\big(\frac{\tau_c}{\tau_{\Omega}}\big)^2} (\hat{e} \times \hat{\omega})\cdot\vec{v}\Bigg]\nn\\
             &&-m\frac{\partial f^0}{\partial E}\frac{\frac{\tau_c}{\tau_{\Omega}}}{4\tau_{\Om}|\hat{e}\times\hat{\om}|\big(1+\big(\frac{\tau_c}{\tau_{\Omega}} \big)^2\big)}\big[-(x+\frac{\tau_c}{\tau_{\Om}}y)\hat{e}\cdot\Vec{v}+(\hat{\om}\cdot\hat{e})(\hat{\om}\cdot\hat{v})(x+\frac{\tau_c}{\tau_{\Om}}y)+(\hat{e}\times\hat{\om})\cdot \Vec{v}(-x\frac{\tau_c}{\tau_{\Om}}+y)\big]\nn\\
              &=& -\frac{\partial f^0}{\partial E} \bigg(\frac{q \tau_c}{1+ \big( \frac{\tau_c} {\tau_{\Omega}} \big)^2} \bigg) \bigg[\delta_{jl}+ \Big( \frac{\tau_c}{\tau_{\Omega}} \Big)^2 \omega_j\omega_l+\Big(\frac{\tau_c}{\tau_{\Omega}}\Big) \ep_{ljk} \omega_k \bigg] \tilde{E}_j v_l\nn\\
              &&-m\frac{\partial f^0}{\partial E}\frac{\frac{\tau_c}{\tau_{\Omega}}}{4\tau_{\Om}|\hat{e}\times\hat{\om}|\big(1+\big(\frac{\tau_c}{\tau_{\Omega}} \big)^2\big)}\Bigg[-\delta_{jl}\Big(x+\Big(\frac{\tau_c}{\tau_{\Omega}}\Big)y\Big)+\om_j\om_l\Big(x+\Big(\frac{\tau_c}{\tau_{\Omega}}\Big)y\Big)+\ep_{ljk}\om_k\Big(-x\Big(\frac{\tau_c}{\tau_{\Omega}}\Big)+y\Big)\Bigg]e_jv_l~.\label{ap7}
\end{eqnarray} 
The second term of RHS of Eq.~(\ref{ap7}) gives a current that is not proportional to the electric field. Hence, it will not contribute to the electrical conductivity components. We may call it a Centrifugal Current($\Vec{j_c}$):
\begin{equation}
  j_{c_{i}}=-mqg \int \frac{d^{3}p}{(2\pi)^3}\frac{\partial f^0}{\partial E}\frac{v^2}{3}\frac{\frac{\tau_c}{\tau_{\Omega}}}{4\tau_{\Om}|\hat{e}\times\hat{\om}|\big(1+\big(\frac{\tau_c}{\tau_{\Omega}} \big)^2\big)}\Bigg[-\delta_{ij}\Big(x+\Big(\frac{\tau_c}{\tau_{\Omega}}\Big)y\Big)+\om_j\om_i\Big(x+\Big(\frac{\tau_c}{\tau_{\Omega}}\Big)y\Big)+\ep_{ijk}\om_k\Big(-x\Big(\frac{\tau_c}{\tau_{\Omega}}\Big)+y\Big)\Bigg]e_j~.\label{ap8}
\end{equation}
\bibliography{non_el}

\end{document}